\documentclass[3p,times]{elsarticle}

\usepackage{ecrc}

\volume{00}

\firstpage{1}

\journalname{Nuclear Physics A}

\runauth{Author1 et al.}

\jid{NUPHA}

\jnltitlelogo{Nuclear Physics A}

\CopyrightLine{2012}{Published by Elsevier Ltd.}

\usepackage{amssymb}
\usepackage[figuresright]{rotating}

\def\npb#1{Nucl.\ Phys.\ B\ {\bf #1}}
\begin{document}

\begin{frontmatter}

\dochead{}



\title{Nucleon resonance production in $K^+N$ interactions}

\author[a]{Bo-Chao Liu}

\address[a]{Department of Applied Physics, Xi'an
Jiaotong University, Xi'an, Shanxi 710049, China}

\begin{abstract}
In this work, we discuss the possibility of studying nucleon resonances in the $K^+N$ interactions.
Based on our model calculations, it is found that the hyperon exchange plays an important role
 for the excitation of nucleon resonances in the $K^+N$ interactions. Therefore, the studies
 on nucleon resonance production in the $K^+N$ interactions can offer us valuable information
 about the couplings of nucleon resonances with strange particles. By selecting suitable decay channels of nucleon resonances, some reactions are also suitable to look for missing resonances.
\end{abstract}


\end{frontmatter}

% \linenumbers

\section{Introduction}
\label{intro}
Understanding the properties and structure of nucleon resonances is a key issue in hadron physics. Up to now,
the status of our knowledge of nucleon resonance is still not satisfying. Besides the parameters of nucleon resonance, such as mass and width, still have large uncertainties, the couplings of $N^*$ with various channels are also not very well known. Currently, our knowledge of nucleon resonance is mainly from the analysis of $N^*$ production processes in $\pi N$ or $\gamma N$ reactions, where the nucleon resonances which have large couplings to $\pi N$ or $\gamma N$ channels are more easily detected. It is believed that lacking measurements with probes different from pion beam is a possible reason for the so-called "missing resonance" problem. In the Particle Data Group (PDG) book\cite{pdg2012}, it is shown that the couplings of nucleon resonances with $\pi N$, $\gamma N$ and $\pi \Delta$ channels are relatively well-known. For other channels, especially for the $K\Lambda$ and $K\Sigma$ channels, our knowledge is still poor. In the PDG summary table for the $N^*$s, only one $N^*$ state is ranked as three stars state for its coupling with $K\Lambda$ or $K\Sigma$ channels, which means "not well determined". For other states, their couplings with strange particles are only ranked as two or one star, which means the knowledge of their couplings is still very poor. On theoretical side, the couplings of $N^*$s with strange particles may give important implications about the internal structures of nucleon resonances. In Ref.\cite{Weise}, the authors argued that the $N^*(1535)$ is a $K\Sigma-K\Lambda$ molecular state. In this picture, large couplings of the $N^*(1535)$ with the $K\Sigma$ and $K\Lambda$ channels are expected\cite{Oset}.
In another work\cite{prl1,prl2}, the authors argued that, if the $N^*(1535)$ has a large coupling to $K\Lambda$ channel, the $N^*(1535)$ may have large five quark component. It is obvious that better knowledge of the couplings of nucleon resonances with strange particles is important for understanding the structure of nucleon resonance and verifying various theoretical models.

In recent years, besides $\pi N$ and $\gamma N$ reactions, some other processes such as $NN$ collision experiments and $J/\Psi$ decay are also used to investigate the property of nucleon resonances. These new experimental data together with $\pi N$ and $\gamma N$ scattering data constitute a better basis for studying nucleon resonances. Some significant improvements have been achieved along this way. With these achievements, one interesting question will be whether there is some other reaction also suitable for studying nucleon resonance. In this work, we want to discuss the possibility of studying nucleon resonance in the $K^+N$ interaction. In fact, in the 1960's and 1970's, there were some experimental studies of nucleon resonance production in the $K^+N$ interactions\cite{data1,data2,data3}. However, with the experimental data with very low statistics at that time, no significant achievements were obtained. On the theoretical side, no systematic studies on this topic have been done either. Our knowledge of the reaction mechanisms of these processes is very poor. Thanks to the pioneering works in the nucleon resonance production in nucleon nucleon collision reactions and the $K^+N$ elastic scattering reactions, it is possible to make some theoretical predictions for the nucleon resonance production in the $K^+N$ interactions. Based on isobar resonance model, we investigate the reaction mechanisms of the
$K^+ N\to K N\eta$ and $K^+p\to K^+K^+\Lambda$ reactions\cite{liu1,liu2}. It is found that the hyperon exchange may play an important role for the excitations of the nucleon resonances in these reactions. Therefore, the process of nucleon resonance production in the $K^+N$ interactions may offer a good place to investigate the coupling of nucleon resonance with strange particles.

\section{Theoretical model and Results}
\begin{figure}[htb]
%%%%%%%%%%%%%
\begin{center}
\includegraphics[scale=0.6]{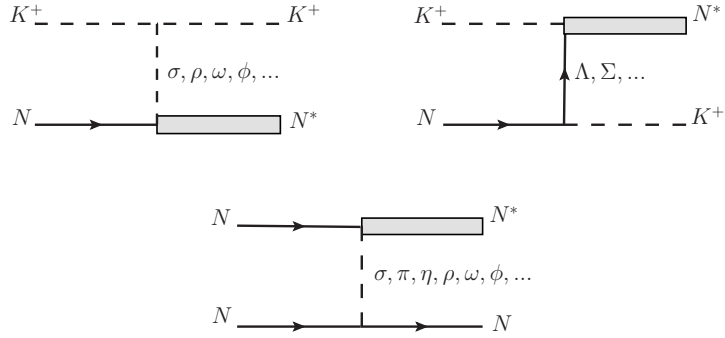}
%%%%%%%%%%%%%
\caption{First row: Feynman diagrams for $K^+N\to K^+ N^*$. Second row: Feynman diagrams for $NN\to NN^*$.}
%%%%%%%%%%%
\label{feynfig1}
\end{center}
\end{figure}
In our work, we adopt the isobar resonance model to study the nucleon resonance production in the $K^+N$ interactions. The interactions among particles are described by the effective Lagrangians. In our model, the nucleon resonance production in the $K^+N$ interactions can be depicted by the Feynman diagrams shown in the first row in Fig.\ref{feynfig1}. If we ignore the decay of the nucleon resonance for this moment, these two diagrams correspond to $t-$channel and $u-$channel diagrams, respectively. Therefore, one may expect that the final kaon will show different pattern of angular distributions if $t-$channel or $u-$channel diagram dominates this reaction. This feature offers the possibility for identifying the reaction mechanism by analyzing the angular distributions of final particles. Here we do not consider the $s-$channel diagram because there is no clear evidence of the existence of pentaquark state. For comparison, we also show the corresponding Feynman diagram for nucleon resonance production in NN collision reaction. In resonance model, both these two processes are dominated by the excitation of nucleon resonance in resonance production energy region. While, in the $K^+N$ interactions, pseudoscalar meson exchange is forbidden and hyperon exchange is allowed. For $NN$ collision reactions, pseudoscalar meson exchange is allowed and hyperon exchange is forbidden. Therefore, the studies of both these two processes can help us know better about the excitation mechanisms of nucleon resonance. In this sense, the studies in these two processes are complementary. As two examples, we present some detailed discussion for the $K^+ N\to KN\eta$ and the $K^+p \to K^+K^+\Lambda$ reactions.

\subsection{$K^+ N\to KN\eta$ near threshold}
For the reaction $K^+ N\to KN\eta$, we consider three different charge channels, i.e. $K^+p\to K^+ p \eta$, $K^+ n\to K^+ n \eta$, and $K^+ n\to K^0 p\eta$. All these three channels can be depicted by the Feynman diagrams shown in Fig.\ref{feynfig1}, where the nucleon and nucleon resonance should be replaced by the corresponding charge states and the nucleon resonance is decaying to the corresponding $N\eta$ channel. It should be noted that, due to the charge conservation law, the allowed exchanging particles between initial states are different for different charge channels. Because we are only interested in the energy region near threshold, we assume that these reactions are dominated by the excitation of $N^*(1535)$ in the intermediate state. This assumption is reasonable because there is no other resonance which gives a significant contribution near threshold\cite{liu1}. In our model, $\sigma$ exchange is also ignored since it was found that the $\sigma$ exchange only plays a minor role for the excitation of $N^*(1535)$ in the reaction $pp\to pp\eta$\cite{nakayama}.

If our assumptions are correct, this reaction may be a good place to investigate the properties of $N^*(1535)$. Next, we will present the results based on the calculations of the corresponding Feynman diagrams shown in Fig. \ref{feynfig1}, which can be obtained by following the standard procedures. For more detailed information about the calculations, please refer to Ref.\cite{liu1}. The main findings of our work are:
\begin{enumerate}
\item
Hyperon exchange may give important contributions in these reactions.

\item
The hyperon exchange and meson exchange induce different pattern of angular distributions of final particles.

\item
The relative roles of individual meson or hyperon exchanges change in different channels, thus a combined analysis of all these reactions may put strong constraints on the model and help us know better about the couplings of $N^*(1535)$ with strange particles.

\end{enumerate}
\subsection{$K^+p \to K^+K^+\Lambda$}
As for the reaction $K^+p \to K^+K^+\Lambda$, we consider the $N^*(1650)$, $N^*(1710)$ and $N^*(1720)$ excitations in the intermediate states. For the corresponding Feynman diagrams, we still use the diagrams shown in the first row in Fig. \ref{feynfig1}, where the nucleon and nucleon resonance represent the corresponding charge states considered in this reaction and the nucleon resonance is decaying to $K^+\Lambda$ channel. In principle, $\Sigma$ and $\phi$ exchanges also contribute. However, because of the very poor knowledge of the couplings of nucleon resonances with the $K\Sigma$ and $N\phi$ channels, we ignore their contributions. Within isobar model, we calculate the corresponding Feynman diagrams(see Ref. \cite{liu2} for more details) and the main results can be summarized as follows:
 \begin{enumerate}
\item
Our results show that the $\Lambda$ exchange gives dominant contributions in lower energies($P_{lab}<$3GeV) and the $\rho$ exchange plays the dominant role in higher energies($P_{lab}>$5GeV).

\item
In the present model, the $N^*(1710)$ gives the dominant contributions in the whole energy region.

\item
Because the $N^*K\Lambda$ vertex can appear twice in the $u-$channel diagram, this reaction is a good place to study the nucleon resonances which have large couplings to the $K\Lambda$ channel and is also suitable to look for missing resonances.

\end{enumerate}
It should be noted that in the present model we reproduce the few available data by adjusting some model parameters. This means that although in the present model we can describe the available data quite well, adjusting the parameters in a reasonable range there is still room for including some other nucleon resonances, for which their couplings to $K\Lambda$ channel are still not well-known. Within our model, it is shown that $N^*(1710)$ plays the most important role in this reaction because of its large coupling to $K\Lambda$ channel. This observation supports the argument that this reaction is possibly a good place to study the nucleon resonance which has large coupling to $K\Lambda$ channel. To my best knowledge, the subprocess $K^+\Lambda\to K^+\Lambda$ of this reaction cannot be observed by experiment directly. So the reaction $K^+p \to K^+K^+\Lambda$ offers a unique place to study the nucleon resonances that have large couplings to $K\Lambda$ channel and to look for missing resonances.

Studying nucleon resonance in the $K^+N$ interactions is also a potentially good way to study the nucleon resonances which have large couplings to $N\rho$, $N\omega$ and $N\phi$ channels. If we consider the vector meson decay channel of the nucleon resonances(see Fig. \ref{feynfig1}), the $N^*$-N-Vector meson vertex can appear twice in the $t-$channel diagram. This feature makes those reactions be suitable for studying nucleon resonances which couple strongly to vector meson channels.

\section{Conclusion}
Based on our model calculations, it is found that the hyperon exchange may play an important role in the nucleon resonance production in the $K^+N$ interactions, which makes the $K^+N$ scattering experiment a potentially good way to study the couplings of nucleon resonances with strange particle channel.
Until now, this topic is only very poorly studied in both theoretical and experimental aspects. To improve our knowledge of the couplings of nucleon resonances with $K\Lambda$ and $K\Sigma$ channels, which is rather poorly known, we think the topic of this work deserves further studies. Furthermore, it will also be interesting to investigate the couplings of nucleon resonances with vector mesons in the $K^+N$ interactions.

\section{Acknowledgments}
This work is supported by the National Natural Science
Foundation of China under Grant No. 10905046 and the Fundamental
Research Funds for the Central Universities.



%% References


%\bibliographystyle{elsarticle-num}



\end{document}